# Plasmonic waveguides with hyperbolic multilayer cladding

Viktoriia E. Babicheva,[1,2] Mikhail Y. Shalaginov,[1] Satoshi Ishii,[1,3] Alexandra Boltasseva[1,4], and Alexander V. Kildishev[1,*]

[1]*School of Electrical and Computer Engineering and Birck Nanotechnology Center, Purdue University, 1205 West State Street, West Lafayette, IN 47907-2057, USA*
[2]*ITMO University, Kronverkskiy, 49, St. Petersburg 197101, Russia*
[3]*International Center for Materials Nanoarchitectonics (MANA), National Institute for Materials Science (NIMS), Tsukuba, Ibaraki 305-0044, Japan*
[4]*DTU Fotonik – Department of Photonics Engineering, Technical University of Denmark, Oersteds Plads 343, Kgs. Lyngby 2800, Denmark*
[*]*kildishev@purdue.edu*

**Abstract:** Engineering plasmonic metamaterials with anisotropic optical dispersion enables us to tailor the properties of metamaterial-based waveguides. We investigate plasmonic waveguides with dielectric cores and multilayer metal-dielectric claddings with hyperbolic dispersion. Without using any homogenization, we calculate the resonant eigenmodes of the finite-width cladding layers, and find agreement with the resonant features in the dispersion of the cladded waveguides. We show that at the resonant widths, the propagating modes of the waveguides are coupled to the cladding eigenmodes and hence, are strongly absorbed. By avoiding the resonant widths in the design of the actual waveguides, the strong absorption can be eliminated.

## 1. Introduction

Metal-dielectric interfaces can support highly confined surface waves known as surface plasmon polaritons (SPPs). SPPs allow one to overcome the diffraction limit, manipulate light at the nanoscale, and merge nanoscale electronics with ultra-fast photonics [1-6]. The capabilities and functionalities of nanophotonic devices can be further improved by utilizing artificially engineered metamaterials with hyperbolic dispersion (so called hyperbolic metamaterials, HMMs) [7-10]. Such materials enable a variety of interesting effects and applications, for example negative refraction [11,12], sub-wavelength imaging [13,14], modified spontaneous emission rate and Purcell effect [15-22], self-induced torque [23], thermal emission engineering [24], complete loss compensation with realistic gain materials [25], improved modulation capabilities [26], volume plasmon polaritons [27-29], as well as designing hyper-crystals [30], tailoring bandgaps [31], vanishing photonic density of states, and isolated nontrivial Dirac cones [32].

Aiming at achieving the highest mode localization and the lowest propagation losses, different waveguide designs were analyzed. One possible approach is to use a hyperbolic material as a guiding medium [33-36]. However, propagation losses in such HMM-waveguides are high [33]. Another approach is to sandwich a dielectric core between claddings with hyperbolic dispersion, forming an HMM-Insulator-HMM (HIH) structure [37]. Here, the term "insulator" is used for convenience and has the same meaning as "dielectric". Previously, it was shown that such designs provide advantages in comparison to conventional plasmonic waveguides [37].

Recently, there was a suggestion to utilize total internal reflection in anisotropic dielectric structures similar to HIH waveguides [38]. While it has been predicted that confinement beyond the diffraction limit in respect to free space wavelength could be achieved, decrease in penetration depth is restricted by the refractive index of constituent dielectric materials. Typical high-index materials, such as silicon and germanium, could only offer a relatively moderate confinement increase factors, namely up to four times at 1.55 μm [38,39]. In contrast, using metals opens up a possibility to confine mode much stronger, in particular, penetration depth can be 10-30 times less than the free space wavelength [37].

In the present work, we investigate the properties of HIH waveguides of finite width with only a few periods of binary metamaterial cells in the claddings (Fig. 1). First, we analyze various HIH waveguides and show their advantages in comparison to metal-insulator-metal (MIM) and insulator-metal-insulator (IMI) waveguides (Section 2). Further, we explore the limitations of the effective medium theory (EMT) by explicitly calculating the properties of metal-dielectric lamellar claddings of the HIH-structure, and discuss the optimum number of the layers (Section 3). These one-dimensional studies are followed by the analysis of more realistic finite-width HIH waveguides and a detailed examination of the feature which emerge in dispersion properties of the structure (Section 4). Finally, we discuss our results and present an outlook in Section 5.

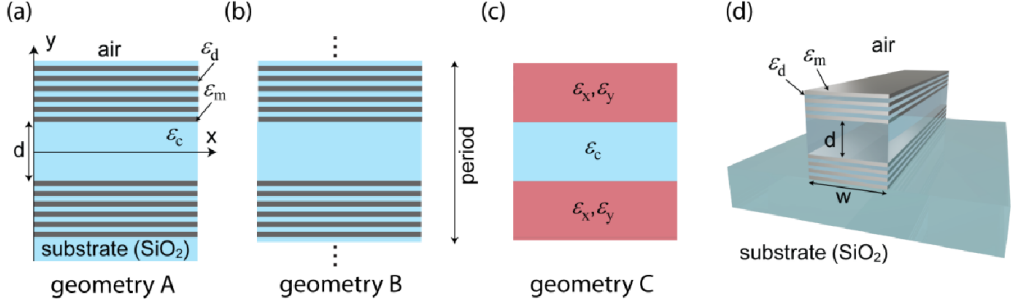

Fig.1. Schematic views of the three geometries: (a) single HIH waveguide (geometry A), $\varepsilon_m$ and $\varepsilon_d$ are the permittivities of metal and dielectric, respectively, $\varepsilon_c$ stands for the permittivity of the waveguide core (b) periodic arrangement of HIH waveguides (geometry B); (c) waveguide with semi-infinite anisotropic effective medium claddings ($\varepsilon_x$, $\varepsilon_y$) that correspond to the lamellar structure (geometry C); (d) finite-width HIH waveguide.

## 2. Comparison with MIM and IMI waveguides

The HIH waveguides considered in the present work consist of planar metal/dielectric lamellar claddings and a dielectric core. Throughout the paper, we assume the metal is silver (permittivity is taken from [40], where the experimental data [41] fitted by the Drude-Lorentz model with five Lorentz oscillators) and the dielectric, including the core and the substrate, is silica (permittivity from [40]). The HMM is modeled as an effective medium (geometry C in Fig.1). By solving the dispersion equation for a particular frequency $\omega$ (corresponding to a free space wavenumber $k_0 = 2\pi/\lambda = \omega/c$) and core thickness $d$, we obtain the complex propagation vector ($k_x$, $k_y$, $k_z$) and calculate other metrics that characterize the waveguide: propagation constant $\beta = \text{Re}[k_z]$, mode index $n_{\text{eff}} = \beta / k_0$, propagation length $L = \text{Im}[2k_z]^{-1}$, penetration depth on the one side of the cladding $\delta = \text{Im}[2k_y]^{-1}$, mode size $D = 2(\delta + d)$, and a figure of merit (FoM) FoM = $L / D$ [42,37]. More details about applying EMT and solving the dispersion equation can be found in [37].

To study the advantages of an HIH waveguide, we have performed a comparative analysis of the HIH characteristics to the characteristics of standard plasmonic waveguide layouts: MIM and IMI. In this study, one of the key parameters is the metal filling fraction $r$ ($r = d_m/(d_m+d_d)$, where $d_m$ and $d_d$ are the thicknesses of the metal and dielectric layers, respectively). Figure 2 shows that at a certain wavelength range and for some parameters, the HIH waveguide outperforms both MIM and IMI. In particular, at $\lambda_t = 1.55$ μm, the FoM of HIH waveguides with $r = 0.15$ is 2.5 times higher than the FoM of MIM waveguides with an identical core thicknesses (Fig. 2(a)). Furthermore, having fixed the propagation length at $L = 592$ μm, the FoM of HIH waveguides with $r = 0.2$ and $d = 30$ nm or $r = 0.16$ and $d = 50$ nm is more than two times higher than the FoM of IMI waveguide with metal thickness $d = 70$ nm (Fig. 2(b)). Thus, HIH structure can provide either better confinement or propagation length in comparison to MIM and IMI designs.

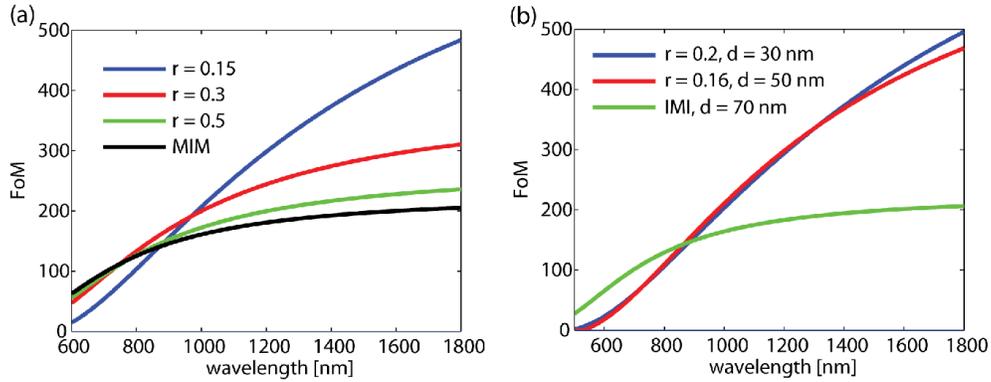

Fig. 2. (a) HIH waveguide FoM for various filling fractions of metal $r$ ($r = d_m/(d_m+d_d) = 0.15$, 0.3, and 0.5) in comparison to FoM of MIM waveguide, in all cases $d = 50$ nm. (b) HIH waveguide FoM for various $r$ and $d$ in comparison to FoM of IMI waveguide. HMM is considered as an effective medium (geometry C in Fig.1).

## 3. HIH waveguide with exact solution

In this section, we perform calculations for the three geometries presented in Fig.1(a)-1(c). In geometry A, the core is symmetrically cladded by metal-dielectric lamellar structure. The number of layers is finite (we consider up to ten) and the bottom and top lamellar structures are bounded by the silica substrate and air respectively. In geometry B, the core and the cladding lamellar structures are the same as in geometry A, but the lamellar-core-lamellar structure is periodic in the y axis. In geometry C, the core is cladded by semi-infinite homogenous hyperbolic media with effective permittivities $\varepsilon_x = r\varepsilon_m + (1-r)\varepsilon_d$ and $\varepsilon_y = \left(r\varepsilon_m^{-1} + (1-r)\varepsilon_d^{-1}\right)^{-1}$. The effective permittivities $\varepsilon_x$ and $\varepsilon_y$ are reasonable approximations for the actual lamellar structures.

The dispersion relation for the geometry C was derived in the previous work [37]. The calculation procedure to derive dispersion relations for geometries A and B based on the T-matrix approach were presented in [43].

In short, wave propagations inside a lamellar structure can be decomposed into two aspects: wave propagation through the interface, from material $l$ to material $l+1$, and wave propagation inside material $l$ of layer thickness $d_l$. The permittivities of the layers $l$ and $l+1$ are $\varepsilon_l$ and $\varepsilon_{l+1}$, respectively. The matrix representation for the former (transmission matrix $D_{l,l+1}$) and the latter (propagation matrix $P_l$) are

$$D_{l,l+1} = \begin{pmatrix} \frac{1}{2}\left(1 + \frac{\varepsilon_{l+1}^2 g_l}{\varepsilon_l^2 g_{l+1}}\right) & \frac{1}{2}\left(1 - \frac{\varepsilon_{l+1}^2 g_l}{\varepsilon_l^2 g_{l+1}}\right) \\ \frac{1}{2}\left(1 - \frac{\varepsilon_{l+1}^2 g_l}{\varepsilon_l^2 g_{l+1}}\right) & \frac{1}{2}\left(1 + \frac{\varepsilon_{l+1}^2 g_l}{\varepsilon_l^2 g_{l+1}}\right) \end{pmatrix} \quad (1)$$

$$P_l = \begin{pmatrix} \exp(ig_l d_l) & 0 \\ 0 & \exp(-ig_l d_l) \end{pmatrix}, \quad (2)$$

respectively, where parameter $g_l = \sqrt{\varepsilon_l k_0^2 - k_z^2}$ [43]. To represent the entire lamellar structure which has $n$ layers, we multiply the transmission matrices and propagation matrices $2n+1$ times.

Thus, geometries A and B are compared with C in terms of the effective indexes and the propagation lengths (Fig. 3). The dependencies of both the effective indexes and propagation lengths vs. wavelength for the considered geometries are very similar. The curves for the geometries A and B are overlapped which essentially means that ten periods is enough for separating single HIH structures such that modes inside different HIH units do not interact with each other. Additionally, they only slightly differ from the geometry C, which validates the use of the EMT method for such calculations.

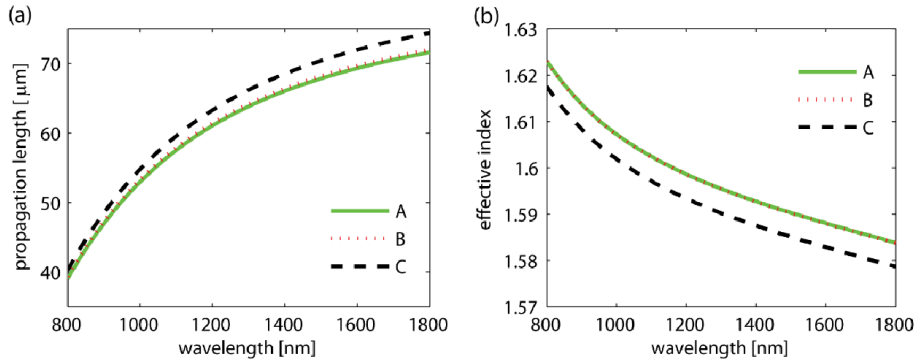

Fig. 3. Dependencies of (a) propagation length and (b) effective index vs. wavelength for the considered waveguide geometries A, B and C, which are shown in Fig.1. The layer thicknesses are 10 nm each, the core thickness is $d = 200$ nm, and the number of periods in the cladding for the geometries A and B is ten.

For a practical realization of the HIH waveguide, it is better to have a minimum possible number of layers in the multilayer cladding. In order to study how the decrease in number of periods affects the waveguide properties, we analyze geometry A (in Fig. 1(a)) and plot the propagation lengths and the effective indexes for the cases of one, two, five, and ten periods (Fig. 4). We also compare the obtained results with the case of semi-infinite EMT cladding. For a cladding consisting of a single pair, the propagation length is two times lower than the EMT-approximated cladding and the index differs by about 3%. Figure Fig. 4 demonstrates that an increase in the number of layers results in a better agreement to the ones obtained using the EMT approximation. In particular, ten pairs are enough to mimic EMT well and to consider the cladding as a hyperbolic medium. Thus, a cladding composed of ten periods of lamellar layers is used for further calculations. One can mention that five layers give close results. By doubling the number of layers accuracy is increased insignificantly, and thus, for the simplified fabrication, one can use five layers as well.

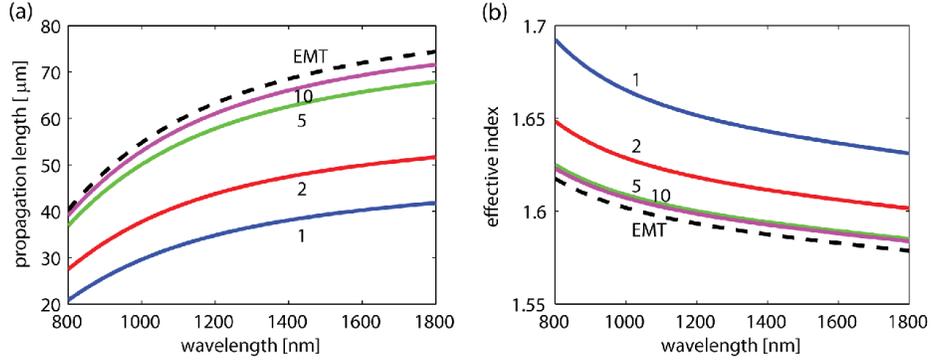

Fig. 4. Comparison of propagation length (a) and effective index (b) dependences vs. wavelength for the geometry A with different number of periods (one, two, five, and ten), as well as for the geometry C, EMT approximation. The other parameters are the same as in Fig. 3.

## 4. Finite-width HIH waveguide

To explore a more realistic device layout, we analyze a finite-width waveguide (Fig.1(d)). The core is cladded with ten periods of identical lamellar structures (10/10-nm-thick silver/silica layers) at the top and bottom, the core thickness is fixed at $d = 200$ nm, and the HIH waveguide is surrounded by air. The eigenmodes of the waveguides with different widths $w$ were found using a commercial finite-element based eigenmode solver (CST Microwave Studio® 2014 [44]). The wavelength is fixed at $\lambda_t = 1.55$ μm. Although the waveguides support a variety of modes with different field distributions (e.g. symmetry, core/cladding intensity distribution ratio, etc.), we focus our study on the modes that are mostly localized inside the core and have a symmetric profile in both x and y directions. A representative field distribution for a waveguide with a width of $w = 1.25$ μm is shown in Fig. 5(a). Numerical simulations can also generate more complicated field profiles inside the core, but we do not study these modes as their excitation in actual devices is difficult.

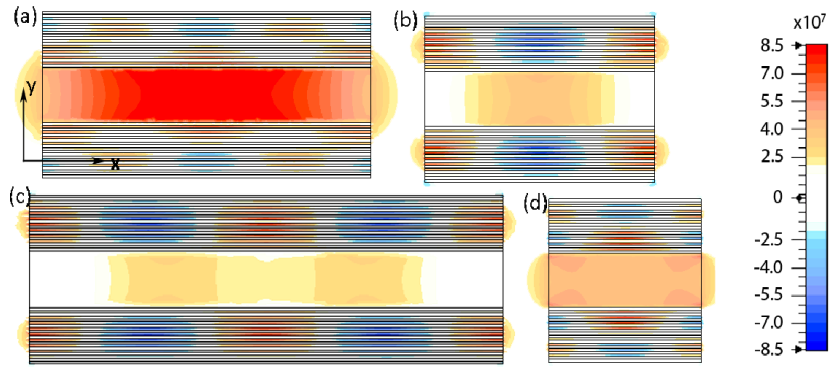

Fig. 5. $E_y$-component field distribution of the modes at $\lambda_t = 1.55$ μm for (a) $w = 1.25$ μm, (b) 0.875 μm, (c) 1.775 μm, and (d) 0.59 μm.

Furthermore, we vary $w$ and analyze propagation length $L$ and propagation constant $β$ of the mode (Fig. 6). At a particular $w$, the propagation length is significantly decreased (see Fig. 6(a)). Some of these dips are accompanied by pronounced resonance profiles of $β$ (see Fig. 6(b)). At non-resonant $w$, values of $β$ are lower than those for the one-dimensional waveguides, or infinite-width, (see increase of $β$ at larger width in Fig. 6(b)). This can be explained by a non-uniform field distribution with lower field penetration into the metal at the edges of waveguide (see Fig. 5(b)), which results in the lower effective index of the mode.

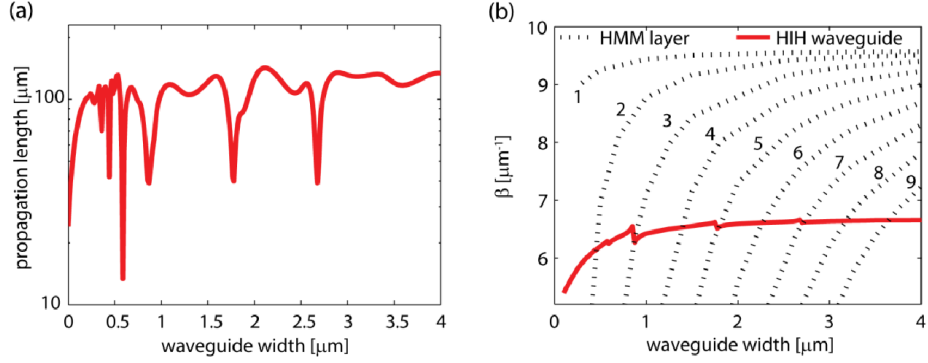

Fig. 6. (a) Propagation length $L$ and (b) propagation constant $\beta$ vs. waveguide width at $\lambda_t = 1.55$ μm.

To determine the origin of the resonances, we analyze field profiles at several different $w$ (Fig. 5(b)-(d)). One can see that $w = 0.875$ μm corresponds to one period of an additional transverse mode in the cladding (see Fig. 5(b)). Doubling the value of $w$ to 1.775 μm corresponds to two periods (see Fig. 5(c)). Cladding modes with more complicated profiles arise for smaller widths $w < 0.6$ μm (see Fig. 5(d)).

For a more detailed analysis, we numerically calculated the eigenmodes of lamellar structures that correspond to a single (top or bottom) cladding. Similar to the whole waveguide analysis, we varied the width of the layer and calculated $\beta$ of the different modes (see Fig. 6(b)). Fig. 7 shows field distributions for the four chosen modes in the 1.5-μm-wide cladding. One can see that resonances arise when the $\beta$ of an odd mode in lamellar structure matches the $\beta$ of the HIH waveguide, i.e. excitation of the odd modes affects the eigenmodes of the HIH waveguide (crossing with lines "3","5", "7" in Fig. 6(b)), while no such effect happens for the even modes (no resonance at crossing with lines "2", "4", "6"). Even cladding modes are not supported in HIH waveguides because of the requirements to have symmetric field distribution inside the waveguide core.

It should be mentioned that including a substrate or other non-uniform surrounding complicates the identification of eigenmodes. Depending on the environment, the eigenmodes of each cladding layer will be different, and the resonances in Fig. 6 will be smeared out or their will be twice as many.

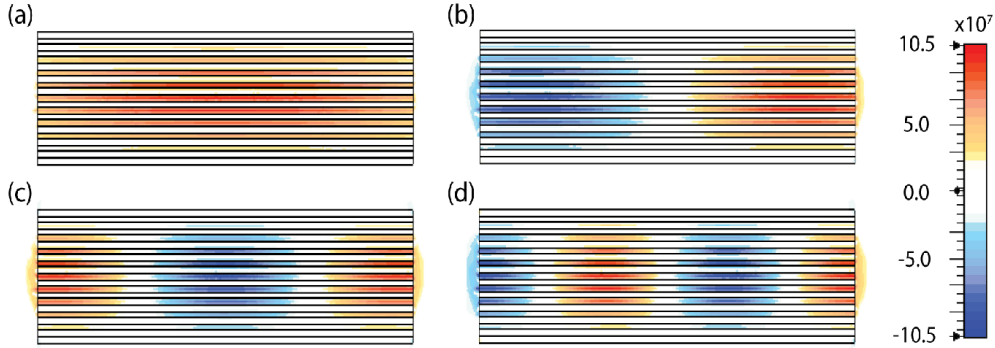

Fig. 7. $E_y$-component field distribution in the multilayer cladding consisting of 10 periods, each layer is 10-nm-thick, $w = 1.5$ μm at $\lambda_t = 1.55$ μm. (a) Corresponds to the mode noted as "1", (b) "2", (c) "3", and (d) "4" in Fig. 6(b).

The dispersion relation for an HMM can be written as

$$\frac{k_x^2 + k_z^2}{\varepsilon_y} + \frac{k_y^2}{\varepsilon_x} = k_0^2, \qquad (3)$$

where $\varepsilon_x$ and $\varepsilon_y$ are the effective medium permittivities (see Section 2).

One can expect that resonances in the cladding correspond to the conditions $k_x w = (2\pi)n_x/2$ and $k_y h = (2\pi)n_y/2$, where $h = 10(d_m + d_d) = 200$ nm is the thickness of the cladding layers and $n_{x,y}$ is an integer. In this simple model, we assume that both $k_x$ and $k_y$ are real values. One can see from Fig. 7, $n_y = 1$ for all resonant patterns. Then, we plot $k_x w/2\pi$ for each mode in Fig. 6(b), i.e., $k_z = \beta + i\alpha$, $k_0 = 2\pi/\lambda_t$, and $k_x$ is obtained from Eq. (3) (Fig. 8). One can see that $k_xw/2\pi = 1$, 2, and 3 coincide with the position of the anomalous spikes in Fig. 6(b) (red triangles in Fig. 8). Since the field penetrates outside the HMM slab in the x-direction (see Fig. 7), the condition $k_xw = \pi n_x$ is not strictly satisfied and it causes small discrepancy between anomalous positions and values of $w = \pi n_x/k_x$ for $n_x$ = 2, 4, and 6 (one can note that $n_x$ does not coincide with mode number).

Thus, the excitation of the cladding modes with resonant transverse profiles and high field localization significantly increases the losses in the finite-width HIH waveguide. On the one hand, these waveguide widths should be avoided while designing practical low-loss HIH structures [2]. On the other hand, lossy plasmonic modes can appear favorable for modulation of propagating modes [45,46], filtering, or waveguide termination [4,47].

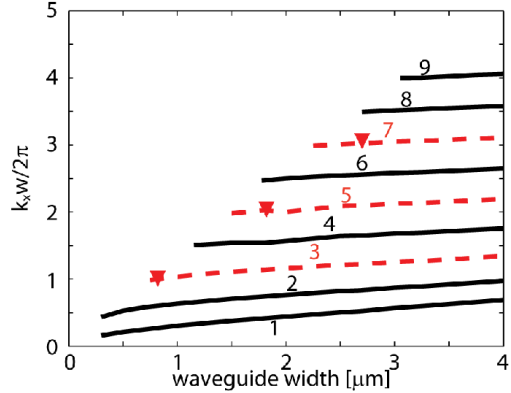

Fig. 8. The number of periods in transverse direction $k_xw/2\pi$ vs. waveguide width at $\lambda_t = 1.55$ μm. Red triangles correspond to the spikes in Fig. 6(b) and well agree with the condition $k_xw = 2\pi$, $4\pi$, and $6\pi$ for the modes 3, 5, and 7, respectively.

## 5. Conclusion

We studied the properties of HIH waveguides and showed that utilizing a multilayer structure with hyperbolic dispersion as a cladding can provide advantages in comparison to standard MIM and IMI plasmonic waveguides, of a similar geometry, in terms of propagation length and mode confinement. We performed the rigorous calculations of the HIH waveguides where multilayer lamellar claddings are treated exactly, without using the effective medium theory. The study showed that when the number of periods exceeds ten, the EMT results closely match the exact calculations, serving as a reasonable approximation. In contrast, rigorous calculations need to be applied for the thinner claddings. We numerically calculated eigenmodes of the HIH waveguide with a finite width and found resonant features for several width values. We showed that the resonant modes inside the cladding of the waveguide with finite width affect the dispersion of waveguides formed by the cladding material. Henceforth, the width of finite width HIH-cladded waveguides should be carefully designed with respect to the resonant modes inside the cladding, which greatly increase the loss of the waveguide, so as to achieve the desired performance.

**Acknowledgments**

The authors would like to thank Nathaniel Kinsey and Clayton DeVault for their kind assistance with manuscript preparation. This work is partially supported by ARO MURI grant 56154-PH-MUR (W911NF-09-1-0539), ARO grant 57981-PH (W911NF-11-1-0359), ONR MURI Grant No. N00014-10-1-0942, and MRSEC NSF grant DMR-1120923 and the Strategic Information and Communications R&D Promotion Programme (SCOPE). V.E.B. acknowledges financial support from SPIE Optics and Photonics Education Scholarship and Kaj og Hermilla Ostenfeld foundation.